\documentclass[11pt]{article}

\usepackage{amsmath,amssymb}
\usepackage{graphicx}
\usepackage{epstopdf}

\newcommand{\be}{\begin{equation}}
\newcommand{\ee}{\end{equation}}
\newcommand{\bea}{\begin{eqnarray}}
\newcommand{\eea}{\end{eqnarray}}
\newcommand{\ba}{\begin{array}}
\newcommand{\ea}{\end{array}}

\newcommand{\AD}[1]{$\ol{\mbox{D~\,}}\!\!\!$#1}

\def \nn {\nonumber}
\newcommand{\eq}[1]{(\ref{#1})}

\newcommand{\tr}{\mbox{tr}}

\newcommand{\ol}{\overline}

\begin{document}


\begin{titlepage}
\begin{flushright}
\end{flushright}

\vfill

\begin{center}
{\large \bf
Holographic QCD with Topologically Charged Domain-Wall/Membranes}
\vfill
{
Feng-Li Lin$^a$\footnote{\tt linfengli@phy.ntnu.edu.tw}
}
 and
{
Shang-Yu Wu$^b$\footnote{\tt loganwu@gmail.com}
}

\bigskip
{\it
$^a$Department of Physics,
National Taiwan Normal University,
Taipei, 116, Taiwan\\
}

and

{\it
$^b$Department of Physics,
National Taiwan University,
Taipei, 106, Taiwan\\
}

\end{center}

\vfill
\begin{abstract}
 We study the thermodynamical phase structures of holographic QCD with nontrivial topologically charged domain-wall/membranes which are originally related to the multiple $\theta$-vacua in the large $N_c$ limit. We realize the topologically charged membranes as the holographic D6-brane fluxes in the Sakai-Sugimoto model. The D6-brane fluxes couple to the probe D8-\AD8 via Chern-Simon term, and act as the source for the baryonic current density of QCD. We find rich phase structures of the dual meson system by varying asymptotic separation of D8 and \AD8. Especially, there can be a thermodynamically favored and stable phase of finite baryonic current density. This provides the supporting evidence for the discovery of the topologically charged membranes found in the lattice QCD calculations. We also find a crossover phase with the limiting baryonic current density and temperature which suggest a Hagedorn-like phase transition of meson dissociation.
\end{abstract}
\vfill

\end{titlepage}

\setcounter{footnote}{0}

\section{Introduction}
  One of the important issues in QCD is to understand its vacuum structure. The non-perturbative vacua such as the ones dominated by the instanton configurations \cite{bpst} are relevant in many aspects of QCD. For example, it was proposed that the instanton is related to the axial U(1) anomaly and yields large $\eta'$ mass \cite{u1a}. However, the effect of dilute instanton gas  is $e^{-N_c}$ and negligible in the large $N_c$ limit. To account for such a discrepancy, a co-dimensional one object of continuous topological charge distribution, namely, the topologically charged membrane \footnote{In the literatures it is usually called topologically charged domain wall. In this paper we prefer to call it topologically charged membrane.} was proposed in \cite{witten79, luscher78}, and later on realized as the D6-brane in the context of AdS/CFT correspondence \cite{witten98}. Interestingly, in parallel with the Wilson loop summarizing the Schwinger effect in the $CP^{n}$ model, L$\ddot{\rm u}$scher  proposed in \cite{luscher78} a Wilson bag term for a 3-form potential $C_3$ over a 3-dimensional hypersurface to characterize the pair production rate of the topological domain wall, i.e.,
 \be\label{luscher-1}
\langle e^{i {e \over \pi} \oint_{\Sigma} C_3}\rangle_{\theta=0} \propto e^{-V_4 \int_0^{2\pi e} d\theta' F(\theta')}
 \ee
where $F(\theta)=-i{1\over 32 \pi^2}\langle \tr F\tilde{F}\rangle_{\theta}$ is the topological charge density with respect to $\theta$-vacuum,  $C_{\mu\nu\rho}=-\tr(A_{\mu}A_{\nu}A_{\rho}+ {3\over 2} A_{[\mu}\partial_{\nu}A_{\rho]})$ is an abelian gauge field of fourth kind such that $4 \epsilon^{\alpha\beta\gamma\delta} \partial_{[\alpha}C_{\beta\gamma\delta]}=\tr F\tilde{F}$.  The 3-dimensional closed hypersurface $\Sigma$ with enclosed 4-volume $V_4$ describes the history of the pair creation and annihilation of the topologically charged membranes. This is the generalization of the Coleman-Schwinger effect for (1+1)-dimensional quantum electrodynamics \cite{Coleman:1976uz}.

 Typically, a 3-form potential is sourced by some charged membranes and will create a long-range Coulomb-type constant force in the ambient (3+1)-dimensional Minkowski space \cite{dvali}. If the force
further couples to light particles, then the Schwinger-like pair production will diminish the 3-form potential and its production rate can be described by \eq{luscher-1}. However, the 3-form potential $C_3$ is not an elementary field but a composite one sourced by the topologically charged membranes, it is not clear how it couples to the pair-produced light fields. Later, we will see that the 3-form potential will couple to the baryonic current from the holographic Chern-Simon term.

   For finite $N_c$, the lattice QCD is reliable and practical to explore the non-pertubative QCD phenomenon. Implementing the lattice chiral symmetry and the associated topological charge density operator, the evidence for the topologically charged membranes in pure-glue $SU(3)$ lattice gauge theory has been reported in \cite{horvath}. Guided by the non-positivity of the two-point function of the topological charge density at non-zero distance \cite{seiler}, which rules out the instanton gas, they found that the dominated configuration in the vacuum is permeated by two oppositely charged sign-coherent connected structures (``sheet"). Each sheet is built from elementary 3D cubes connected through 2D surface.

    On the other hand, it is believed that this topologically charged membrane  is related to closely juxtaposed dipole-like  D6-\AD6-branes in the holographic QCD model \cite{witten98,thacker} without considering its effect on the hadron dynamics.  The holographic QCD model has been generalized by Sakai-Sugimoto \cite{ss1,ss2} to include the hadron physics by using the D4-D8-\AD8 brane configuration to capture the quark dynamics. This model is based on Witten's model for 4-dimensional pure Yang-Mills theory by wrapping $N_c$ D4-branes on a Scherk-Schwarz circle \cite{wittenMQCD}, and then putting additional probe $N_f$ D8 and \AD8-branes transverse to the circle.  The open strings connecting D4 and D8 or \AD8 then provide the chiral fermions as quarks in the fundamental representation of both the gauge group $U(N_c)$ and the flavor group $U(N_f)_L \times U(N_f)_R$.  In the strongly coupled regime in the large $N_c$ limit, the D4-branes are condensed into Witten's geometry \cite{wittenMQCD}, which is of cigar-shape  so that the D8 and \AD8 are curved and smoothly connected into U-shaped configuration with an asymptotic separation $L$ at infinity.  The worldvolume theory of the probe D8-brane in Witten's geometry successfully realizes the meson and hadron physics \cite{ss1,ss2}.   This U-shaped D8-\AD8 configuration geometrically realizes the spontaneous chiral symmetry breaking at low energy. Besides that, the Sakai-Sugimoto model has also been generalized to the deconfined phase \cite{Aharony:2006da}.

   Motivated by the appearance of the topologically charged membrane in the lattice simulation \cite{horvath} and its implication of L$\ddot{\rm u}$scher's Wilson bag, it is interesting to study this effect in the various phases of the Sakai-Sugimoto model. In this paper, we will realize the topologically charged membranes by the Ramond-Ramond (RR) 8-form fluxes, which can be understood as condensate of the smeared D6-\AD6 branes.  Interestingly, from the Chern-Simon term of the probe D8-\AD8 branes, we find that the RR 8-form will couple to the spatial component of the U(1) part of the flavor gauge field, which is holographic dual to the coupling between L$\ddot{\rm u}$scher potential $C_3$ and the baryonic current. Moreover, realizing the external baryonic fields as the D6-branes' flux  helps us to understand its back reaction to the D8-\AD8 probe branes' shape, and remove the singular chemical potential configuration. This makes possible the exploration of the full parameter space of the phase diagram of Sakai-Sugimoto model with nontrivial vacuum dominated by topologically charged membranes.  Similar story has happened in introducing the finite baryon number density in Sakai-Sugimoto model \cite{baryond}, our investigation for relation between the topologically charged membrane and the baryonic current will follow the same line.

   Recently, the baryonic current coupled to the external baryonic electric field in Sakai-Sugimoto model has been considered in \cite{baryonic}, where the baryonic Ohm's law for the decay of the baryonic electric field is identified and can be understood as the Schwinger effect. Instead, we find that there is no imaginary part in D8's DBI action, and the topologically charged membranes realized as the D6-branes' fluxes are  dynamically stable. Therefore,  the Schwinger effect as suggested in \cite{luscher78} or  \cite{witten98} through the nucleation of the baryonic current is absent for topologically charged membrane. However, we find very rich phase structures of thermodynamics.

   The paper is organized as following. In the next section we give configuration of the source D6-brane fluxes as the holographic dual of the topologically charged membranes in QCD. Especially our configuration will couple to boundary baryonic current density and yield non-trivial thermodynamics. In section 3 the cusp configuration of probe D8-\AD8 in the confined phase is solved, its holographic thermodynamic is derived and the phase structures are displayed. We find that there are rich phase structures as we vary the asymptotic separation of D8 and \AD8. For example, we find a crossover phase with a limiting baryonic current density indicating a Hagedorn-like phase transition of meson dissociation, i.e. quark deconfinement. More important, there exists stable topologically charged membranes which support the lattice QCD's results in \cite{horvath}.  Similar considerations for the high temperature phase of holographic QCD are done in section 4 in which there are more rich phase structures due to the additional temperature dependence. Finally we conclude our paper in section 5.

\section{The topologically charged membranes in Sakai-Sugimoto model}
The holographic dual of 4-dimensional, large $N_c$ QCD with massless flavors was proposed in \cite{ss1,ss2}. This model is constructed by embedding $N_f$ D8-\AD8 probe branes in the D4-brane background. In addition, we will also turn on the condensed $N_6$ D6-branes as the topologically charged membranes.  The brane setting is arranged as following:
\begin{eqnarray}
\begin{array}{ccccccccccc}
& 0 & 1 & 2 & 3 & (4) & 5 & 6 & 7 & 8 & 9 \\
N_c \;\mbox{D4} & \times & \times & \times & \times & \times &&&&& \\
N_f\; \mbox{D8-\AD8}
& \times & \times & \times & \times &  & \times & \times & \times &\times & \times \\
N_6\; \mbox{D6}
& \times & \times & \times &  &  &   & \times & \times &\times & \times \nn\\
\end{array}
\label{D4D8D6}
\end{eqnarray}
The dual QCD lives on $R^4$ spanned by the coordinates $x^0$ to $x^3$ where the above D6s look like the co-dimension one domain wall/membranes.
For simplicity, we will only consider the topologically charged membrane with $x^3$ as its transverse direction. Furthermore, we will smear the D6-brane fluxes along the $x^3$-direction uniformly, which represents an uniform topologically charged membrane distribution.  In the above brane setting, the D4 and D6-branes are condensed into geometric background but with the D8-\AD8 as the probe brane whose worldvolume theory will capture the essence of QCD's meson physics \cite{ss1}. Moreover, we assume
\be\label{limitrel}
N_6 \ll N_f \ll N_c,
\ee
to suppress D6s' back-reaction to the D4 background and also its contribution to the vacuum energy of the  dual meson's (thermo-)dynamics of D8-\AD8.

   As the supergravity background, the above D6-brane fluxes satisfy the source equation\footnote{In this paper, we consider field equations in the string frame.}
\begin{eqnarray}\label{f8eom}
\partial_{U}(\sqrt{-g}F^{U0126789})=J_6
\end{eqnarray}
where $U$ is the radial coordinate of Witten's geometry. The explicit form of the D6-fluxes $F_8$ is determined by the D6-brane source $J_6$. To have an energy-scale-independent thermodynamics of the holographic QCD, it turns out that we have to introduce the D6-fluxes in the following form
\be\label{f8ex}
F_8= N_6\; \delta(U-U_c) dx^0\wedge dx^1\wedge dx^2 \wedge dU \wedge d\Omega_4
\ee
where the D6-branes are uniformly smeared along the $x_3$-direction\footnote{A smooth $U$-profile D6-fluxes may introduce inhomogeneous current density profile along $U$-direction, this is beyond the scope of simple thermodynamics considered in this paper.}.  We then plug the $F_8$ of \eq{f8ex} into source equation \eq{f8eom} to deduce the source $J_6$. From the term
$\partial_U \delta(U-U_c)$ we can see the leading terms of  $J_6$ contain both positive and negative singular charge density at $U=U_c$. This implies a closely packed D6-\AD6 pair located at $U=U_c$. This is consistent with the lattice result in \cite{horvath} in which appears a vacuum with two oppositely charged sign-coherent connected structures (``sheet").

  Before we continue, we should make sure the D6-fluxes will not cause large back reaction to the D4-branes' background metric.  To check this, we only need to compare the on-shell actions of the D4 and D6-fluxes.
The ratio of D6 to D4's kinetic terms (in string frame) is
\be\label{ncsup}
{F^2_8 \over F^2_4 }\sim {N_6^2 \over N_c^2} [\delta(U-U_c)]^2
\ee
where the 4-form fluxes $F_4$ is sourced by the $N_c$ D4-branes, and we have omitted the irrelevant power of $U$ factor.
With the assumption \eq{limitrel}, the D6s' kinetic term is $1/N^2_c$ suppressed if we firstly regularize the singular profile \eq{f8ex} and then take large $N_c$ limit. Moreover, in this way the D6-fluxes will not add the vacuum energy to the (thermo-)dynamics of the dual meson system of D8-\AD8.

  Given \eq{f8ex} we can obtain a Hodge dual 2-form $\hat{F}_{34}$, which is associated with an axion $\hat{a}_4$ by $\hat{F}_{34}:=\partial_3 \hat{a}_4 - \partial_4 \hat{a}_3$. The axion couples to the topological charge density of dual QCD
\be\label{ax1}
\int_{S\times R^4} \hat{a}_4\; \tr F \tilde{F}.
\ee
At the same time, the $\theta$ parameter of dual QCD vacuum can be defined as
\be\label{baryo-axion}
\theta=\lim_{U\rightarrow \infty}\int dx^4 \hat{a}_4=\int^{\infty}_{U_{\rm KK}} dU\int dx^4 \hat{F}_{U4}
\ee
which we have used the Stokes theorem. Note that this yields zero value of $\theta$ for our D6-fluxes since $\hat{F}_{U4}=0$ by construction.  This is in contrast to the case considered in \cite{witten98,Bergman:2006xn} in which a D6's 2-form $\hat{F}_{U4}=n_6/U^4$ corresponding to $J_6=0$ in \eq{f8eom} is used, so that the $\theta$ parameter is nonzero and finite.

  In Sakai-Sugimoto model, the meson dynamics is described by the DBI action of the probe D8-\AD8-branes with $U(N_f)$ gauge fields as the holographical dual of the meson fields. Especially, the diagonal $U(1)$ part of the flavor group is holographically related to the baryonic current $\bar{\psi}^a \gamma^{\mu} \psi_a$ \cite{baryonic,baryond}. Besides, we should also include the Chern-Simon term
\be\label{scs}
S_{CS}=T_{8}\int{C_{p+1}\wedge \tr(2\pi\alpha'F)}^{(8-p)/2}
\ee
if there are non-trivial RR-form fields $C_{p+1}$ generated by Dp-brane sources.  In the presence of above D6-branes acting as the topologically charged membranes in the holographic dual QCD, the source Chern-Simon term to be incorporated is
\begin{equation}\label{cs-d6}
S_{CS}= T_{8}\int{C_{7}\wedge \tr(2\pi\alpha'F)}=N_f T_{8}\int{F_{8}\wedge a_3 dx^3}
\end{equation}
where $a_{3}\equiv \frac{2\pi\alpha'A_{3}}{\sqrt{2N_{f}}}$ is the $x^3$-component of the diagonal U(1) part of the gauge field.  As usual, $a_{\mu}$ can be considered as the chemical potential of the baryonic current. On the other hand, if we identify $\hat{a}_4$ in \eq{ax1} as the baryo-axion\footnote{Compactifying $C_7$ in \eq{cs-d6} on the internal 4-sphere of the Witten's geometry of D4-branes, the reduced 3-form potential can be thought as the 4-dimensional Hodge dual of the baryo-axion, that is the field sourced by the topologically charged membrane.}  proposed in \cite{dvali}, i.e., $\hat{F}_{34}=*_{10}F_8$, then the suggestive form of \eq{cs-d6} provides a holographic realization of the conjectured coupling proposed in \cite{dvali} between the baryo-axion and the baryonic current.  In some sense, the spontaneous baryonic current induced by the topologically charged membranes is analogous to the one by the skyrmions.

\section{Confined phase with topologically charged membranes}
 In this section, we will discuss the thermodynamics of Sakai-Sugimoto model in the confined phase with topologically charged membranes discussed in the previous section. The main idea of the Sakai-Sugimoto model is to realize the meson dynamics of strongly-coupled QCD by probe D8-\AD8 branes in the Witten's geometry of $N_c$  non-extremal D4-branes, which is holographically dual to strongly coupled large $N_c$ Yang-Mills theory in the confined phase.

The Witten's geometry of the D4-brane background describing QCD confined phase is given by
\begin{eqnarray}\label{ssmetric}
&& ds^{2}=\left(\frac{U}{R}\right)^{\frac{3}{2}}\left(\eta_{\mu\nu} dx^{\mu}dx^{\nu}+f(U)(dx^4)^2 \right)+\left(\frac{R}{U}\right)^{\frac{3}{2}}\left(\frac{dU^{2}}{f(U)}+U^{2}d\Omega^{2}_{4}\right)\\
&& F_{4}=\frac{2\pi N_{c}}{\Omega_4}\epsilon_{4},\;\; e^{\phi}=g_{s}\left(\frac{U}{R}\right)^{\frac{3}{4}},\;\;  f(U)=1-\frac{U^3_{\rm KK}}{U^3}
\end{eqnarray}
where $\mu,\nu=0,1,2,3$, and $\Omega_4=8\pi^2/3$ and $\epsilon_4$ are the volume and volume form of the internal unit $S^4$.    The parameter $R$ is related to the string coupling and string length by $R^3=\pi g_s N_c l_s^3$, and the metric describes a throat geometry with the $x^4$ circle smoothly shrink to a tip at the IR endpoint $U=U_{\rm KK}$ by requiring the period of $x^4$ to be $4\pi R^{3/2}/3U^{1/2}_{\rm KK}\equiv 2\pi/M_{\rm KK}$ to avoid conical singularity. Moreover, the relations between the gravity and the gauge theory quantities are
\begin{eqnarray}
R^3 = \frac{g_{\rm YM}^2 N_c l_s^2}{2 M_{\rm KK}}, \quad
 U_{\rm KK} = \frac29 g_{\rm YM}^2 N_c M_{\rm KK} l_s^2,\quad
g_s = \frac{g_{\rm YM}^2}{2 \pi M_{\rm KK}l_s}
\label{grel}
\end{eqnarray}
where $g_{\rm YM}$ is the gauge coupling of the dual 4-dimensional Yang-Mills theory.

 The $N_f$ probe D8-\AD8-branes are embedded in such a way with $x_4$-direction as the only transverse direction, this yields the induced metric on its worldvolume as follows
\begin{equation}
ds^{2}_{D8}=\left(\frac{U}{R}\right)^{\frac{3}{2}}\eta_{\mu\nu} dx^{\mu}dx^{\nu}+\left(\frac{R}{U}\right)^{\frac{3}{2}}\left[\left(\left(\frac{U}{R}\right)^{3}{f(U)} (\partial_U x^4)^2+\frac{1}{f(U)}\right)dU^{2}+U^{2}d\Omega^{2}_{4}\right].
\end{equation}

  The D8-\AD8 worldvolume dynamics is described by the usual DBI action with $U(N_f)$ gauge field as follows
\begin{equation}\label{sdbi}
S_{DBI}=-N_f T_{8}\int{d^{9}x e^{-\phi}\sqrt{-\det(g+2\pi \alpha'F)}}=-\texttt{N}\int{dU U^{4}\sqrt{f(U)(\partial_U x^4)^2+\left(\frac{R}{U}\right)^{3}({1\over f(U)}+(\partial_U a_{3})^2)}}
\end{equation}
where we have defined
$\texttt{N}\equiv \frac{N_{f}T_{8}\Omega_{4} V_{3}\beta}{g_{s}}$ where $V_3=\int dx^1dx^2dx^3$ and $\beta=\int dx^0$, and the $x^3$-component of the diagonal U(1) part of the gauge field $a_{3}\equiv \frac{2\pi\alpha'A_{3}}{\sqrt{2N_{f}}}$.

  From the dictionary between the bulk and boundary quantities, for example the ones given in \cite{baryond} for the discussions of the finite holographic baryon number density,  the boundary value of the above bulk background gauge field $a_3$ corresponds to the chemical potential of the $x^3$-component of the baryonic current of the dual QCD. The dual baryonic current density can also be identified from the normalizable mode of  $a_3$.

  Moreover, we should also add the Chern-Simon term \eq{scs} for the nonzero RR 8-from \eq{f8ex} corresponding to the topologically charged membranes in the QCD vacuum. This yields
\be\label{csaction}
S_{CS}=\texttt{N} \; n_b \int dU \delta(U-U_c) a_3(U)
\ee
which acts as a source term for the boundary baryonic current. In the above, we cut off the volume of the $x^3$-direction to $L_3$, and define the baryonic current density of the dual QCD   by
\be
n_b\equiv {N_f T_8 N_6 L_3 \over \texttt{N}}={3 \over 8\pi^2} {g_s N_6 \over  V_2 \beta}, \qquad V_2\equiv\int dx^1 dx^2.
\ee

\subsection*{Solving D8-\AD8 configuration with a cusp}

   The Euclidean on-shell action  of the probe D8-brane will encode the thermodynamics of the holographic QCD. To acquire the information, we need to solve the D8-brane configuration.

\begin{figure}[t]
\begin{center}
\includegraphics[width=12cm]{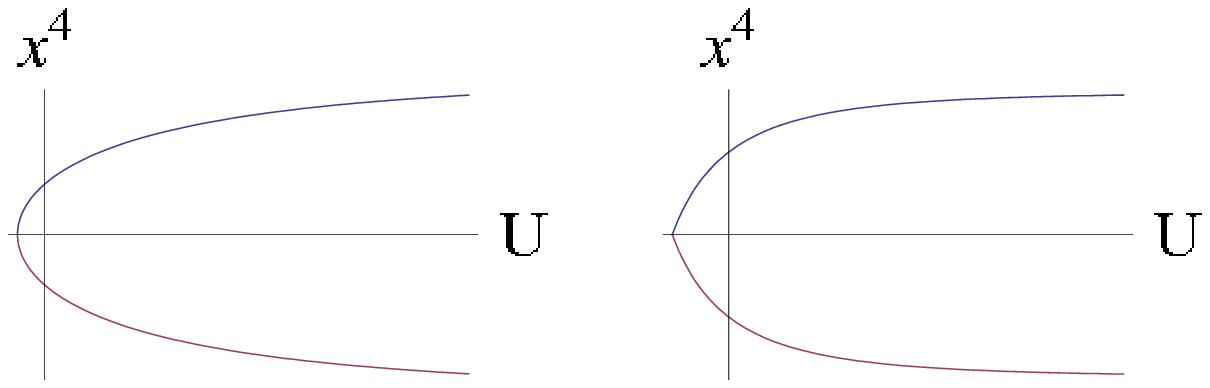}
\caption{\qquad Smooth D8-\AD8 \qquad \qquad \qquad \qquad D8-\AD8 with a cusp \qquad}
\label{cusp}
\end{center}
\end{figure}

   In  flat space-time, the D8-brane will be straight and flat. However, due to the cigar shape of Witten's geometry, the probe D8-brane will force to  bend  into \AD8-brane at  $U_0$. Integrating \eq{x4os} to get $x^4(U)$ with fixed asymptotic separation of D8 and \AD8, one will get a U-shaped D8-\AD8-brane configuration, which presents the chiral symmetry breaking of the dual QCD in a geometric way \cite{ss1}. Moreover, once the D6-brane fluxes are turned as \eq{f8ex}, the above smooth U-shaped D8-\AD8 will then develop a cusp as shown in Fig.\ref{cusp} due to the pulling of the force from the D6-brane source, which signifies the existence of a new thermodynamical phase \footnote{This is similar to the case of finite baryon number density, for example see the last reference in \cite{baryond}.}.

    We first solve the gauge field $a_3$ from its equation of motion derived from the total action of D8-\AD8, which  is
\be\label{eoma3}
 \Pi_{a_3} =\texttt{N} \; n_b \; \Theta(U-U_c), \qquad U_c > U_{\rm KK}
\ee
where the properly normalized conjugate momentum to $a_3$ is defined by
\be
\Pi_{a_3}\equiv  {\delta S_{DBI} \over \delta (\partial_U a_3)}=-\texttt{N}{ U R^3 \partial_U a_3 \over \sqrt{f(U)(\partial_U x_{4})^2+\left(\frac{R}{U}\right)^{3}({1\over f(U)}+(\partial_U a_{3})^2)}}, \label{pia3}
\ee
and the step function on the right-handed-side comes from the delta function D6-fluxes source encoded in \eq{csaction}.

 From \eq{pia3} we can get
\begin{eqnarray}
\partial_U a_3=
 -\frac{n_b}{UR^3}\sqrt{\frac{f(U) (\partial_U x^4)^2+(\frac{R}{U})^3{1\over f(U)}}{1-\frac{n_b^2}{U^5R^3}}},  \label{pa3}
\end{eqnarray}
The above choice of the overall  sign is to be consistent with \eq{eoma3} and \eq{pia3}.

     On the other hand, there is no source term for the equation of motion of $x^4$, we instead need introduce an integration constant $C$. With the help of \eq{pa3}, we have
\be\label{x4os}
(\partial_U x^4)^2={({R\over U})^3  C \over (f(U))^2(H(U) -C)}
\ee
with
\be
H(U)\equiv U^{8}f(U)\left(1-\frac{n_b^2}{U^5R^3}\right).
\ee
The integration constant is assigned by relating it to the proper cusp angle
\be
\tan \theta_c \equiv \sqrt{g_{44}\over g_{UU}}\partial_U x^4\mid_{U=U_c}
\ee
and it yields
\be
C \equiv U^8_c f(U_c)\left(1-\frac{n_b^2}{U_c^5R^3}\right)\sin^2\theta_c.
\ee

   To further integrate out \eq{x4os} to obtain the full D8-\AD8 configuration we need one more condition, which is to fix the asymptotic separation between D8 and \AD8
\be\label{Lfix}
L = 2\int^{\infty}_{U_c}{dU \; \partial_U x^{4}}.
\ee

 Moreover, the size of the proper cusp angle $\theta_c$ can be determined by the force balance condition between the D8-\AD8's tension $f_{D8}$ and the pulling of D6 $f_{D6}$, i.e.,
\be\label{fbala}
f_{D8}+f_{D6}=0.
\ee

 The proper tension of D8-\AD8 is obtained by varying its Hamiltonian density with respect to the proper distance $U_{c}$ along the D8 brane, that is,
\begin{eqnarray}
f_{D8}&=&\frac{1}{\sqrt{g_{UU}}}\frac{\delta H_{D8}}{\delta U}\mid_{U=U_{c}}\\
&=&\frac{1}{\sqrt{g_{UU}}}\left[-\texttt{N}{\cal H}(U_{c})+\int^{\infty}_{U_{c}}{\frac{\delta H_{D8}}{\delta
(\partial_U x^4)}\frac{\partial (\partial_U x^4)}{\partial U}} \mid_{U=U_c,\; with\; n_{b},L \;fixed}\right] \label{f82}
\end{eqnarray}
where the Hamiltonian density is the Legendre transformation of the DBI action \eq{sdbi} as
\be
H_{D8} = - S_{DBI}+  \int dU\; \Pi_{a_3} \partial_U a_3
=\texttt{N}\int_{U_c}^{\infty} dU\; {\cal H}\left(x^4(U)\right)
\label{ham-c}
\ee
with
\be\label{ham-d}
{\cal H}\left(x^4(U)\right)\equiv  U^4 \sqrt{\left(f(U) (\partial_U x^4)^2+\left(\frac{R}{U}\right)^3{1\over f(U)}  \right)\left(1-\frac{n_b^2}{R^{3}U^{5}}\right)}.
\ee
  The resultant proper D8-\AD8 tension is \footnote{ While varying $\partial_U x^4$ with $U_c$ we should keep the asymptotic separation \eq{Lfix} fixed, this yields
\be
 \partial_U x^4(U_c)=\int_{U_c}^{\infty} \frac{\partial (\partial_U x^4)}{\partial U} \mid_{U=U_{c}, with\; n_{b},L \; fixed} \; dU
 \ee
 from which we can evaluate the second term in \eq{f82} along with the Hamilton's equation  $\partial_U \left(\frac{\delta H_{D8}}{\delta (\partial_U x^4)}\right)=0$.}
\be\label{fd8f}
f_{D8}=-\texttt{N}R^{3\over 4}U_c^{13\over 4} \sqrt{1-\frac{n_b^{2}}{U_c^{5}R^3}}\cos\theta_c.
\ee

   To obtain the pulling force of the D6-fluxes, we need to know the on-shell Chern-Simon action which can be obtained by plugging \eq{x4os} into \eq{pa3}, and then integrating it over $U$ to get $a_3(U)$. Resultantly,
\be\label{scsos}
S_{CS}|_{on-shell}= \texttt{N} n_b  a_3(U_c)= \texttt{N}  \left[a_3(\infty) n_b + \int_{U_c}^{\infty} dU {n_b^2\left( {U \over R}\right)^{3\over 2}\over \sqrt{H(U)-C}} \right].
\ee
Therefore, the force by D6 flux is
\be\label{fd6f}
f_{D6}=-{1 \over \sqrt{g_{UU}}}{\delta S_{CS} \over \delta U}\mid_{U=U_c, \; with\; n_b, L \; fixed}=-\texttt{N}^2{n_b^2 \over f_{D8}} \left({U_c \over R}\right)^{3\over 2} .
\ee

  Given the explicit form of $f_{D8}$ and $f_{D6}$, the force balance condition \eq{fbala} yields
\be\label{costc}
\cos^2\theta_c={\frac{n_b^2}{U_c^5R^3}\over 1-\frac{n_b^2}{U_c^5R^3}}.
\ee

\subsection*{Absence of dynamical instability via tunneling}

  Though the factors $\sqrt{H(U)-C}$ and $\sqrt{1-\frac{n_b^2}{U^5R^3}}$  appearing in the on-shell action seem implying the possibility of dynamical instability via tunneling such as the Schwinger effect if these factors are turned into complex. This usually happens for a DBI action with the electric field larger than its maximal field strength. However, recall that we in fact turn on not electric but magnetic field $\partial_U a_3$ for the holographic baryonic current density. Thus, we will not expect such a dynamical instability in our case. This is also in contrast to the case considered in \cite{baryond} for the finite holographic baryon number density, in which a electric field $F_{U0}$ is turned on in the DBI action, so there is a possible dynamical instability for the Lorentzian configurations. Indeed, \eq{costc} implies $\frac{n_b^2}{U_c^5R^3}<1/2$  so that $\cos^2\theta_c <1$ for a physical cusp configuration, thus the above factors are real for $U>U_c$ and the dynamical instability is absent. This implies a maximal  baryonic current density $n_{b}<\sqrt{\frac{U^{5}_{c}R^{3}}{2}}$ beyond which the cusp configuration cannot be sustained,  and it could imply the meson dissociation.

\subsection*{Thermodynamics of holographic QCD}

  From the usual thermodynamical and holographical interpretation of the Euclidean path integral in the grand canonical ensemble at temperature $T$
\be
\sum_n e^{-(E_n-\mu Q_n)/T}=e^{-\Omega(T,\mu)/T}:=e^{-(E-\mu n_b-T S)/T},
\ee
 we can identify the on-shell Euclidean total action as the Gibbs free energy density {\it with fixed chemical potential} $\mu$, i.e.,  \be\label{totosa}
\Omega(\mu; n_b)= S^{E}_{DBI}|_{on-shell}+S^{E}_{CS}|_{on-shell}.
\ee
The holographic chemical potential of the baryonic current in dual QCD is identified as as the boundary value of gauge field, that is,
\be
\mu=\texttt{N} a_3(\infty)
\ee
and the on-shell Euclidean Chern-Simon action differs from \eq{scsos} by an overall minus sign, together with the on-shell Euclidean DBI action obtained  from \eq{sdbi}, \eq{pa3} and \eq{x4os}
\be\label{sdbios}
  S_{DBI}^E|_{on-shell}=\texttt{N} \int^{\infty}_{U_{c}}{dU\frac{R^{3\over 2} U^{13\over 2}}{\sqrt{H(U)-C}}}.
\ee

    Instead, we can also work in the canonical ensemble {\it with fixed charge density} $n_b$ by performing the Legendre transformation to obtain the Helmholtz free energy density
\be
F(T,n_b)=E-TS=\Omega+ \mu n_b
\ee
and from \eq{scsos} and \eq{sdbios} we obtain
\be\label{freeeos}
F(n_b)=\texttt{N} \int_{U_c}^{\infty} dU \frac{R^{3\over 2} U^{13\over 2}\left( 1-\frac{n_b^{2}}{U^{5}R^3} \right)}{\sqrt{H(U)-C}}
\ee
which is nothing but the on-shell value of the Hamiltonian density \eq{ham-c} and \eq{ham-d} as expected from the definition of the Helmholtz free energy \footnote{In the confined phase, the temperature dependence is trivial and its thermodynamics will be thought as the zero temperature case.}. Moreover, the relation between chemical potential and charge density in the canonical ensemble can be obtained via\footnote{Or, we can vary the Hamiltonian density \eq{ham-c} so that
\be
\mu(n_b)= \frac{\delta H_{D8}}{\delta U}\mid_{U=U_{c}} \cdot {d U_c \over d n_b}+{\partial H_{D8} \over \partial n_b}= \texttt{N} \sqrt{g_{UU}(U_c)}f_{D8}{d U_c \over d n_b}+{\partial H_{D8} \over \partial n_b},\nn
\ee and obtain ${d U_c \over  d n_b}$ by varying the condition \eq{Lfix} over $n_b$ so that
\be
{d U_c \over  d n_b}={\int_{U_c}^{\infty} dU {\partial (\partial_U x^4) \over \partial n_b} \over \partial_U x^4\mid_{U=U_c} - \int_{U_c}^{\infty} dU {\partial (\partial_U x^4) \over \partial  U_c}}.\nn
\ee
 This is useful for numerical implementation. }
\be\label{cmu-1}
\mu(n_b) =  {d F\over d n_b}.
\ee
 Inverting this relation we can get $\Omega(\mu)=\Omega(\mu, n_b(\mu))$.  Furthermore, the thermodynamical stability is characterized by the baryonic current susceptibility
 \be
 \chi={d^2 F\over dn_b^2}.
 \ee
If $\chi$ is positive, it is stable; otherwise, it is not.

\subsection*{Phase structures}

   Since the condition \eq{Lfix} of fixed asymptotic separation is nontrivial so that it is impossible to obtain the analytical form for the above thermodynamical quantities, instead we will substitute the proper cusp angle by \eq{costc} and numerically solve $U_c(n_b)$ from \eq{Lfix} and obtain the thermodynamical quantities.  Moreover, we may expect rich phase structures due to the factors such as $\sqrt{H(U)-C}$ in the thermodynamical quantities. This is in contrast to the simple phase structure considered in \cite{baryond} for the case with non-zero holographic baryon number density. In fact, the most peculiar we have found is that the thermodynamical behaviors strongly depend on the value of the asymptotic separation $L$. As $L$ is varied, there are the following phases: (a) thermodynamically unfavored and unstable phase,  (b) thermodynamically favored and stable one, (c) thermodynamically unfavored but stable one, and (d) crossover ones between the above phases.  The feature of the rich phase structures is quite similar to the one of holographic entanglement entropy characterized by a critical asymptotic separation \cite{Klebanov:2007ws}.

   In our numerical computations we have set \footnote{We should caution the readers here. In this chosen set of units, the 't Hooft coupling is not much larger than one so that the SUGRA approximation is not  good. However, it is conventionally used for the numerical computations in the literatures \cite{baryonic, baryond} with surprisingly good results in comparison with experimental value. }
\be
U_{\rm KK}=0.5, \qquad R=1 \Rightarrow L\le \frac{2\pi}{3}\sqrt{\frac{R^{3}}{U_{\rm KK}}}\simeq2.96.
\ee
Moreover, the thermodynamics for the confined phase for a given $L$ can be just summarized in two plots, one is $\mu$ v.s. $n_b$, the other is $\Delta \Omega$ v.s. $\mu$, from which we can obtain $\Delta F(n_b)$.  The $\Delta \Omega$  is the difference of the Gibbs free energy between topologically charged membrane state ($n_b\ne 0$) and vacuum state ($n_b=0$). Similarly for $\Delta F(n_b)$. It is also interesting to see how the tip of cusp changes while varying $n_b$ from $U_c(n_b)$, however, we omit it in the following because it is not relevant for thermodynamics.

\subsubsection*{Thermodynamically unfavored and unstable phase}
\begin{figure}[b]
\begin{center}
\begin{minipage}{7.5cm}
\begin{center}
\includegraphics[width=6cm]{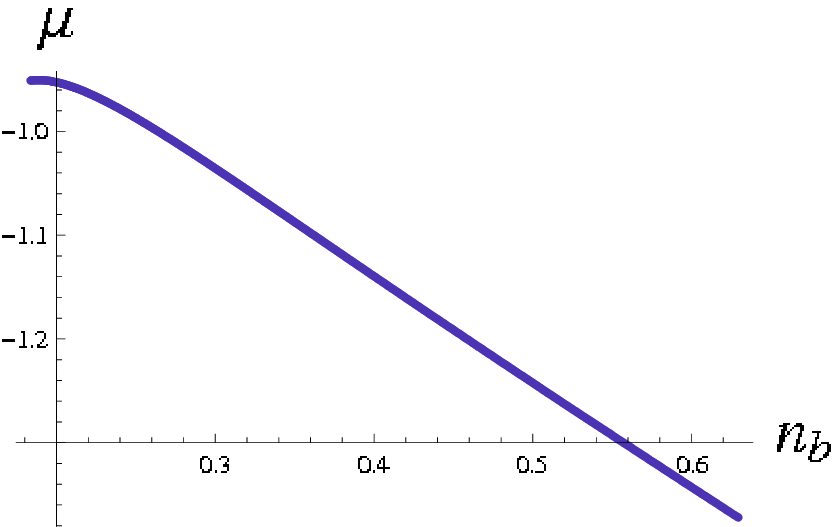}
\caption{$\mu$ v.s. $n_b$ ($L=0.4$)}
\label{munb04}
\end{center}
\end{minipage}
\hspace{4ex}
\begin{minipage}{7.5cm}
\begin{center}\hspace*{-4ex}
\includegraphics[width=6cm]{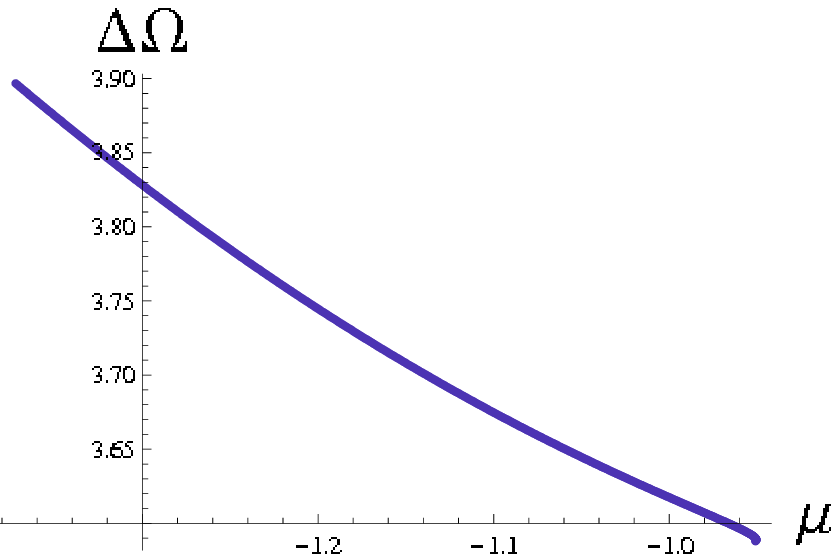}
\caption{$\Delta \Omega$ v.s. $\mu$ ($L=0.4$)}
\label{Omegamu04}
\end{center}
\end{minipage}
\end{center}
\end{figure}

The results of a typical case with $L=0.4$ for such a phase are summarized in Fig.\ref{munb04}-\ref{Omegamu04}. We can see that:
\begin{itemize}
\item  $\Delta \Omega$ in Fig.\ref{Omegamu04} is positive so that the phase is thermodynamically unfavored.

\item Moreover, this phase is also thermodynamically unstable as can be seen from the negative susceptibility derived from the slope of Fig.\ref{munb04}.  This is also suggested by the negativity of the chemical potential $\mu$.
\item  In these plots, only the part with the baryonic current density $n_b>0.2$ is shown. In our numerical results not shown in Fig.\ref{munb04}-\ref{Omegamu04}, we find that for very small $n_b$ the Gibbs free energy is far higher and is thermodynamically unfavored. Therefore we just show the parameter space with smaller $\Delta\Omega$ region which is also more realizable without numerical convergence problem.   Similar situation happens for the other cases of different $L$'s.
\end{itemize}

\subsubsection*{Crossover phase}

\begin{figure}[t]
\begin{center}
\begin{minipage}{7.5cm}
\begin{center}
\includegraphics[width=6cm]{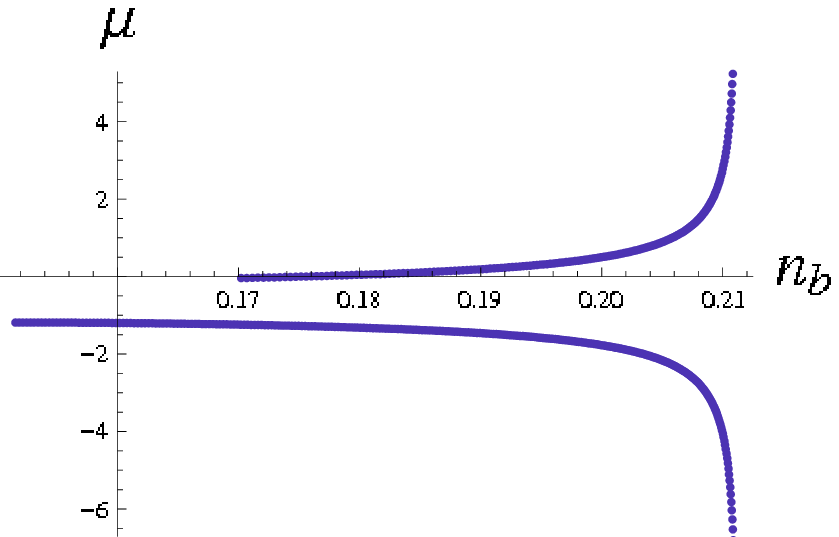}
\caption{$\mu$ v.s. $n_b$ ($L=0.65$)}
\label{munb65}
\end{center}
\end{minipage}
\hspace{4ex}
\begin{minipage}{7.5cm}
\begin{center}\hspace*{-4ex}
\includegraphics[width=6cm]{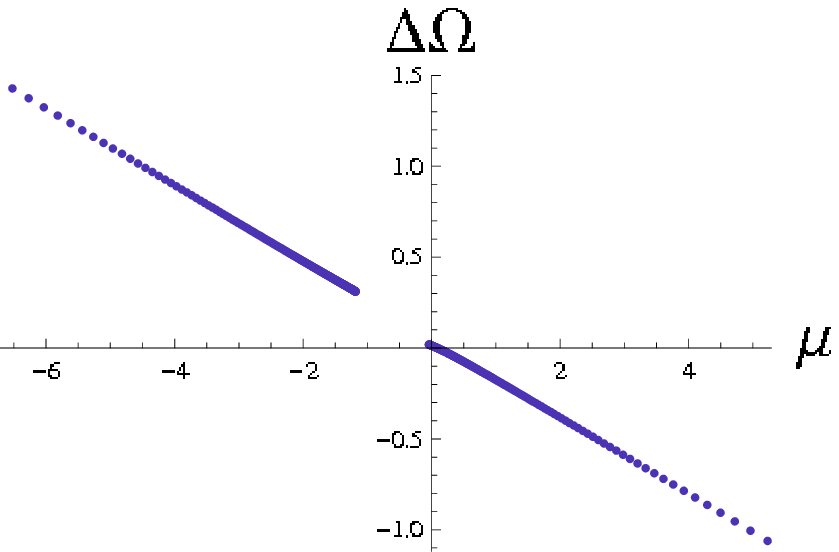}
\caption{$\Delta \Omega$ v.s. $\mu$ ($L=0.65$)}
\label{Omegamu65}
\end{center}
\end{minipage}
\end{center}
\end{figure}

 In between the thermodynamically unstable phase for $L<0.6$ and the stable ones for $L>0.7$, there is a crossover as shown in
Fig.\ref{munb65}-\ref{Omegamu65} for $L=0.65$. We see that
\begin{itemize}

 \item Two branches appear in Fig.\ref{munb65} because there are two values of chemical potential for a fixed baryonic current density.

 \item One branch is still thermodynamically unfavored and unstable, but the other is both thermodynamically favored and stable as can be seen from Fig.\ref{munb65} and \ref{Omegamu65}.

 \item From Fig.\ref{munb65} we see there exists a maximal value of $n_b$ at which the both chemical potential and susceptibility are divergent, and thus limiting. This is similar to the limiting susceptibility for the Hagedorn phase transition \cite{Hagedorn transition, Lin:2007gi}.  However, this limiting baryonic current density is smaller than the one required by $\cos^2\theta_c<1$ to sustain the cusp configuration. This suggests that the cusp configuration cannot be sustained for the smaller $n_b$ than expected, and the limiting behavior might be a signature for meson dissociation which bears Hagedorn-like behavior as expected.  This could be related to the similar Hagedorn behavior for the holographic entanglement entropy considered in \cite{Klebanov:2007ws}.

   Therefore, our result shows that the topologically charged membranes can induce quark deconfinement even at zero temperature.

 \item Another interesting feature for this phase is the sign change of the chemical potential for baryonic current density as we vary $L$.  This is different from the usual crossover in QCD of finite baryon number density, for which there is no sign change of chemical potential for baryon number density \cite{QCD Phase diagram}.  Instead, it reminds us the BCS-BEC crossover in condensed matter system \cite{BCS-BEC crossover} where there is a sign change of chemical potential from the BCS phase ($\mu>0$, weakly coupled) to the BEC phase ($\mu<0$, strongly coupled)  as the number density increases.

\end{itemize}

\subsubsection*{Thermodynamically favored and stable phase}

\begin{figure}[t]
\begin{center}
\begin{minipage}{7.5cm}
\begin{center}
\includegraphics[width=6cm]{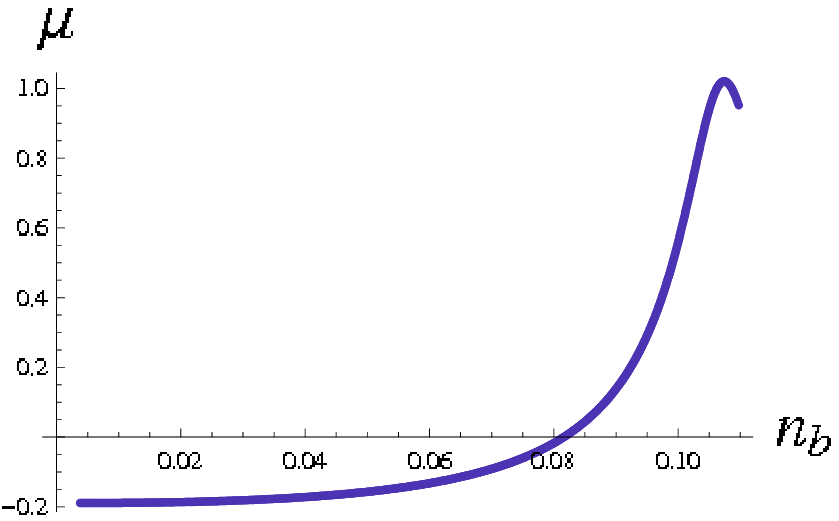}
\caption{$\mu$ v.s. $n_b$ ($L=1$)}
\label{munb10}
\end{center}
\end{minipage}
\hspace{4ex}
\begin{minipage}{7.5cm}
\begin{center}\hspace*{-4ex}
\includegraphics[width=6cm]{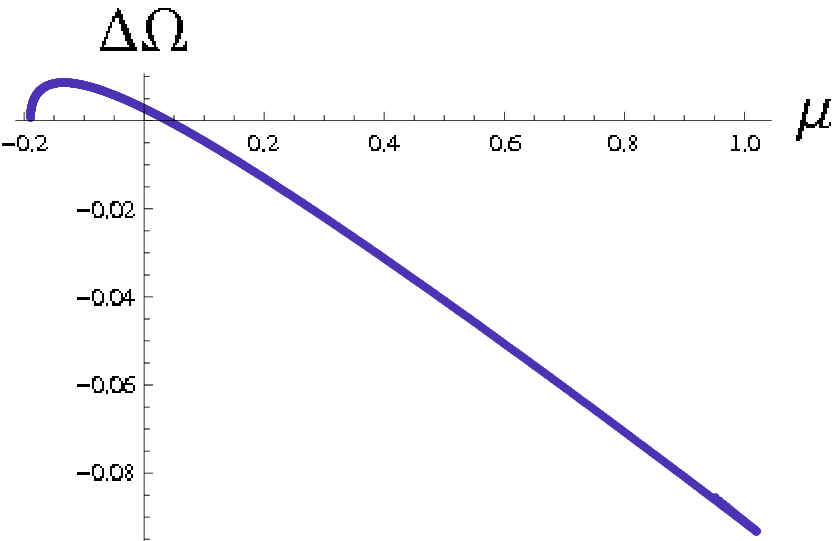}
\caption{$\Delta \Omega$ v.s. $\mu$ ($L=1$)}
\label{Omegamu10}
\end{center}
\end{minipage}
\end{center}
\end{figure}

   One of the goal of this work is to search for the new stable vacuum of large $N_c$ QCD with non-trivial topologically charged membrane structure. The results of a typical case with $L=1$ for such a phase are shown in Fig.\ref{munb10}-\ref{Omegamu10}.

 \begin{itemize}
\item  $\Delta \Omega$ in Fig.\ref{Omegamu10} is negative for the most of positive $\mu$ region, it implies this phase
is thermodynamically favored.  It is interesting to see there is some unfavored region for negative $\mu$ as the tail of the crossover.

\item Moreover, this phase is also thermodynamically stable as can been from the positive susceptibility derived from the slope of Fig.\ref{munb10}. Similarly, there is a small region of negative susceptibility as the tail of the crossover.

\item     Therefore, there exists a thermodynamically favored and stable phase with non-zero baryonic current density in the confined phase of holographic QCD. This provides the supporting evidence to the finding of the topologically charged membranes in the lattice QCD calculations reported in \cite{horvath}.

\end{itemize}

\subsubsection*{Thermodynamically unfavored but stable phase}

\begin{figure}[t]
\begin{center}
\begin{minipage}{7.5cm}
\begin{center}
\includegraphics[width=6cm]{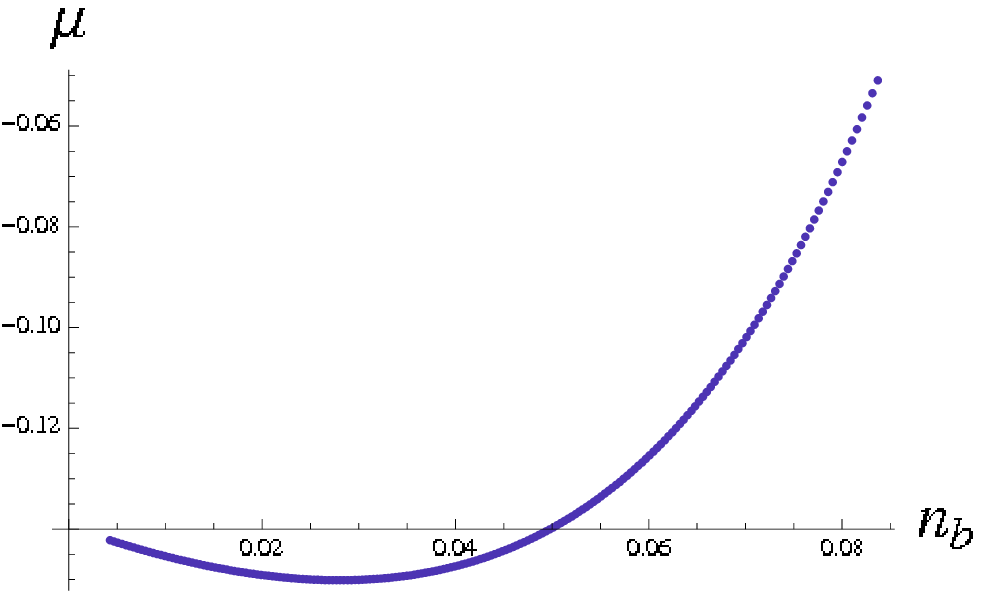}
\caption{$\mu$ v.s. $n_b$ ($L=1.5$)}
\label{munb15}
\end{center}
\end{minipage}
\hspace{4ex}
\begin{minipage}{7.5cm}
\begin{center}\hspace*{-4ex}
\includegraphics[width=6cm]{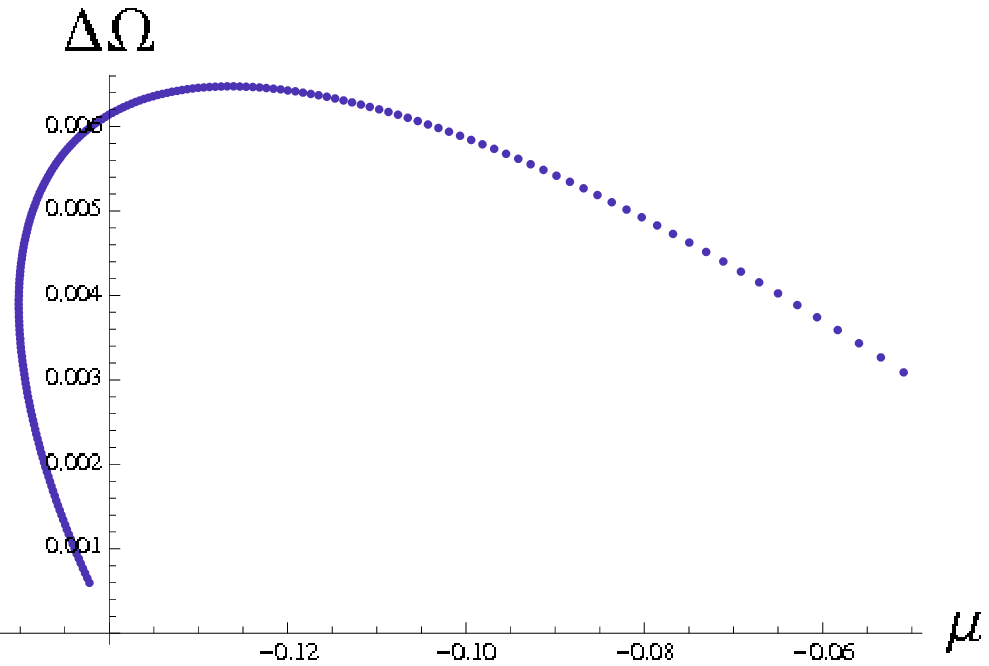}
\caption{$\Delta \Omega$ v.s. $\mu$ ($L=1.5$)}
\label{Omegamu15}
\end{center}
\end{minipage}
\end{center}
\end{figure}

   As $L$ goes higher, the positive $\Delta \Omega$ part will gradually dominate the negative one, and finally we will end up at such a phase where the negative $\Delta \Omega$ part will disappear but it is still thermodynamically stable. The typical results with $L=1.5$ are given in Fig.\ref{munb15}-\ref{Omegamu15}, and this phase persists for larger $L$ \footnote{Our numerical calculation is reliable up to $L\approx 2.2$, for higher $L$ we have serious convergence problem since $U_c$ is very close to $U_{\rm KK}$ and it becomes harder to solve \eq{Lfix} numerically.  }. It is straightforward to read off the thermodynamical behavior as before.  Since $\Delta \Omega$ is small, it suggests that the topologically charged membranes may have quite a chance to sustain in the vacuum even it is relatively unfavored.

In summary, as the asymptotic separation $L$ of D8-\AD8 increases, the meson system of the holographic QCD goes from a thermodynamically unfavored and unstable phase through a crossover, and then becomes thermodynamically favored and stable. Finally, it ends at a thermodynamically unfavored but stable phase. Such a rich phase structure at zero-temperature system is expected for a strongly interacting theory as holographic QCD. This is in contrast to the free theory with trivial thermodynamical phase at zero temperature.

\section{High temperature phase with topologically charged membranes}

  The background geometry dual to high temperature phase of QCD is obtained from the metric \eq{ssmetric} by wrapping the Euclidean time and $x^4$ cycles as in the usual Witten's cigar like geometry for BTZ black hole \cite{wittenMQCD,Lin:2007gi}. This was considered first in \cite{Aharony:2006da} and the resultant (Euclidean) metric is \footnote{Dilaton's and 4-form flux's profiles are the same as in the confined phase.}
\begin{equation}\label{htmetric}
ds^{2}=\left(\frac{U}{R}\right)^{\frac{3}{2}}\left(f(U)d\tau^{2}+\delta_{ij}dx^{i}dx^{j}+dx^{2}_{4}\right)+\left(\frac{R}{U}\right)^{\frac{3}{2}}\left(\frac{dU^{2}}{f(U)}+U^{2}d\Omega^{2}_{4}\right)
\end{equation}
where $f(U)=1-\frac{U^3_{\rm T}}{U^3}$, and the period of the Euclidean time direction is determined to be
\be\label{holot}
{4\pi \over 3}{\sqrt{R^{3}\over U_{\rm T}}} \equiv \beta={1\over T}
\ee
 which is identified as the inverse temperature of the dual QCD.
As for the confined phase,  the induced metric on the probe D8-\AD8 in the background \eq{htmetric} is
\begin{equation}
ds^{2}_{D8}=\left(\frac{U}{R}\right)^{\frac{3}{2}}\left(f(U)d\tau^{2}+\delta_{ij}dx^{i}dx^{j} \right)+\left(\frac{R}{U}\right)^{\frac{3}{2}}\left[\left(\left(\frac{U}{R}\right)^{3}  (\partial_U x^4)^2+\frac{1}{f(U)}\right)dU^{2}+U^{2}d\Omega^{2}_{4}\right].
\end{equation}
 It then results in the DBI action \footnote{The Lorentzian DBI and Chern-Simon actions are considered until we consider thermodynamics.} of D8-\AD8
\be
S_{DBI}=-\texttt{N}\int{dU U^{4}\sqrt{f(U)(\partial_U x^4)^2+\left(\frac{R}{U}\right)^{3}\left(1+f(U) (\partial_U a_{3})^2\right))}}
\ee
where $\texttt{N}$ and $a_3$ are defined as for the confined phase.  However, the Chern-Simon action is the same as \eq{csaction}.

     Since the $U$-direction is now no longer caped as in the confined phase but the $\tau$-direction is, it is then possible for the D8-\AD8 to disjoin as temperature goes high enough. This indicates the chiral symmetry will restore in the high temperature \cite{Aharony:2006da}. With this mind we solve the equations of motion.

    The equation of motion for $a_3$ is
\be
\partial_U a_3 = -\frac{n_b}{UR^3}\sqrt{\frac{(\partial_U x^4)^2+\left(\frac{R}{U}\right)^3{1\over f(U)}}{f(U)-\frac{n_b^2}{U^5R^3}}},
\ee
 and the one for $x^4$ is
\be
(\partial_U x^4)^2=\frac{(\frac{R}{U})^3 f^{-1} C_T}{G(U)-C_T}
\ee
with
\be
G(U)\equiv U^8\left(f(U)-{n_b^2 \over U^5 R^3}\right)
\ee
and the integration constant $C_T$ is related to the proper cusp angle $\theta_c$ as follows
\be
C_T\equiv U_c^8\left(f(U_c)-{n_b^2 \over U_c^5 R^3}\right)\sin^2\theta_c.
\ee
For the chiral symmetry breaking phase the proper cusp angle $\theta_c$ is nonzero and will be determined as in the confined phase. For the chiral symmetry restoration phase, $\theta_c=0$ so that $\partial_U x^4=C_T=0$ which describes the parallel D8 and \AD8-branes. Formally, we can consider both cases in a unified way, and should be carefully choose $\theta_c$ and the corresponding quantities when we tune the temperature or $U_T$ in numerical calculations.

  The Hamiltonian density derived from the DBI action by the Legendre transformation is
\be\label{ham-ht}
H_{D8}=\texttt{N}\int_{U_c}^{\infty} dU\;  U^4 \sqrt{\left( (\partial_U x^4)^2+\left(\frac{R}{U}\right)^3{1\over f(U)}  \right)\left(f(U) -\frac{n_b^2}{U^5R^3}\right)}
\ee
from which we can obtain the proper D8-\AD8 tension for the cusp development, that is
\be
f_{D8}=-\texttt{N}R^{3\over 4}U_c^{13\over 4} \sqrt{f(U_c)-\frac{n_b^{2}}{U_c^{5}R^3}}\cos\theta_c.
\ee

Similarly we can derive the pulling of D6-fluxes from the Chern-Simon action, and arrive the following force balance condition for the cusp (in the chiral symmetry breaking phase) which yields
\be
 \cos^2\theta_c={\frac{n_b^2}{U_c^5R^3}\over f(U_c)-\frac{n_b^2}{U_c^5R^3}}.
\ee
Along with the condition for the fixed asymptotic separation of  D8 and \AD8, i.e. \eq{Lfix}, we can solve $U_c(n_b)$  as in the confined phase. Note that requiring $\cos^2\theta_c\le 1$, the current density is bounded as $n_b\le U^{\frac{5}{2}}_{c} R^{3\over 2} \sqrt{\frac{1-(\frac{U_{T}}{U_{c}})^{3}}{2}}$ for a given $T$, or a bounded temperature as $T\le \frac{3}{4\pi}(U^{3}_{c}-\frac{2n^{2}_{b}}{U^{2}_{c}})^{1\over 6}$ for a given $n_b$.

      As for the consideration of the thermodynamics of the dual QCD in the grand canonical ensemble, the Gibbs free energy is the on-shell Euclidean total action and we obtain
\be
\Omega(T, \mu; n_b) = - \mu n_b + F(T,n_b)
\ee
where the chemical potential $\mu=\texttt{N} a_3(\infty)$ as before, and the free energy density
\be
F(T,n_b)=\texttt{N} \int_{U_c}^{\infty} dU \frac{R^{3\over 2} U^{13\over 2}\left( f(U) -\frac{n_b^{2}}{U^{5}R^3} \right)}{\sqrt{f(U)\left(G(U)-C_T^2\right)}}
\ee
is the same as the on-shell value of Hamiltonian density \eq{ham-ht} as expected; however, the nontrivial temperature dependence is implicit in the factor $f(U)=1-\frac{U^3_{\rm T}}{U^3}$ which is related to $T$ via \eq{holot}. Also, one can obtain $\mu(T,n_b)$ through
\be
\mu(T,n_b)={d F(T,n_b) \over d n_b}.
\ee

\subsection*{Phase structures}
At high temperature there are three phases: cusp, parallel (chiral symmetry restored phase), and vacuum ($n_{b}=0$). To determine which phase is preferred, we have to compare their Gibbs free energies. We denote the Gibbs free energy for cusp phase as $\Omega_{c}$, for parallel one as $\Omega_{p}$ and for vacuum as $\Omega_{v}$, and their differences as $\Delta\Omega_{cv}=\Omega_{c}-\Omega_{v}$, $\Delta\Omega_{cp}=\Omega_{c}-\Omega_{p}$, and $\Delta\Omega_{pv}=\Omega_{p}-\Omega_{v}$, respectively. Also, the chemical potentials for the cusp and parallel phase are denoted as $\mu_{c}$ and $\mu_{p}$ respectively.

   In order to perform the numerical computation, it is easier to fix either the temperature $T$ or the current density $n_B$ as what we will do in the following.  We will set $R=1$ as in the confined phase.

\subsection*{Fixed $T$}

  With $U_T$ fixed, the equations of motion are formally similar to the confined phase, we may expect similar thermodynamical properties as for the confined phase. Indeed, this is the case and there exists a thermodynamically favored and stable phase in the configuration space. Therefore, for conciseness, we will not show all the plots but only the relevant ones, and summarize the main features as following:

\begin{figure}[t]
\begin{center}
\begin{minipage}{7.5cm}
\begin{center}
\includegraphics[width=6cm]{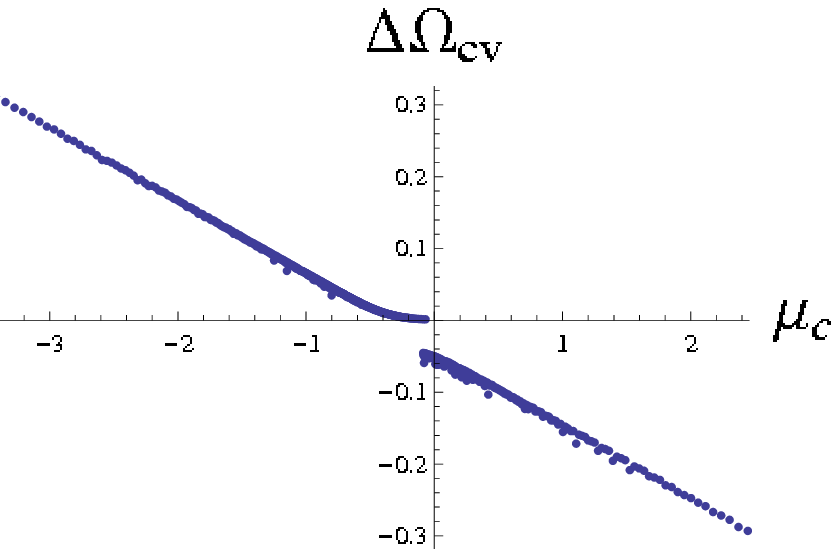}
\caption{$\Delta \Omega_{cv}$ v.s. $\mu_c$ ($L=0.7, T\simeq0.169$)}
\label{Omegacvmuc07_UT05}
\end{center}
\end{minipage}
\hspace{4ex}
\begin{minipage}{7.5cm}
\begin{center}\hspace*{-4ex}
\includegraphics[width=6cm]{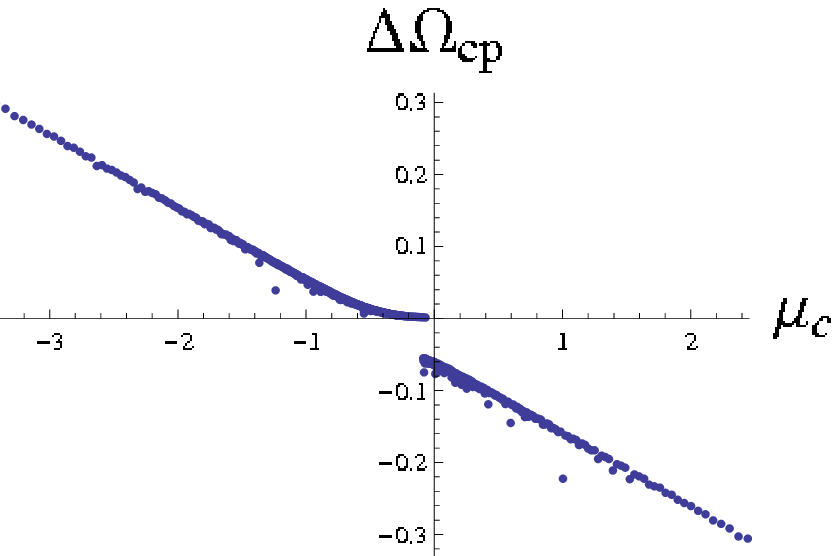}
\caption{$\Delta \Omega_{cp}$ v.s. $\mu_c$ ($L=0.7, T\simeq0.169$)}
\label{Omegacpmuc07_UT05}
\end{center}
\end{minipage}
\end{center}
\end{figure}

\begin{figure}[t]
\begin{center}
\begin{minipage}{7.5cm}
\begin{center}
\includegraphics[width=6cm]{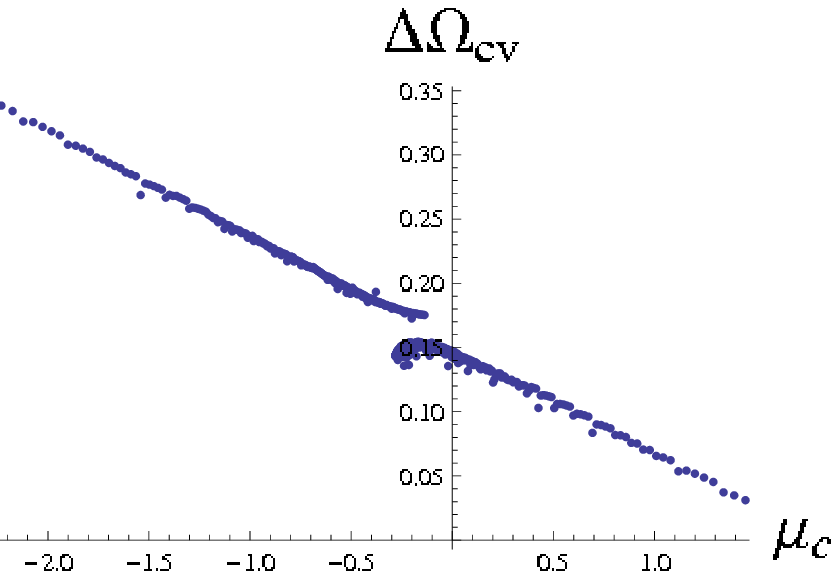}
\caption{$\Delta \Omega_{cv}$ v.s. $\mu_c$ ($L=0.7, T\simeq0.207$)}
\label{Omegacvmuc07_UT075}
\end{center}
\end{minipage}
\hspace{4ex}
\begin{minipage}{7.5cm}
\begin{center}\hspace*{-4ex}
\includegraphics[width=6cm]{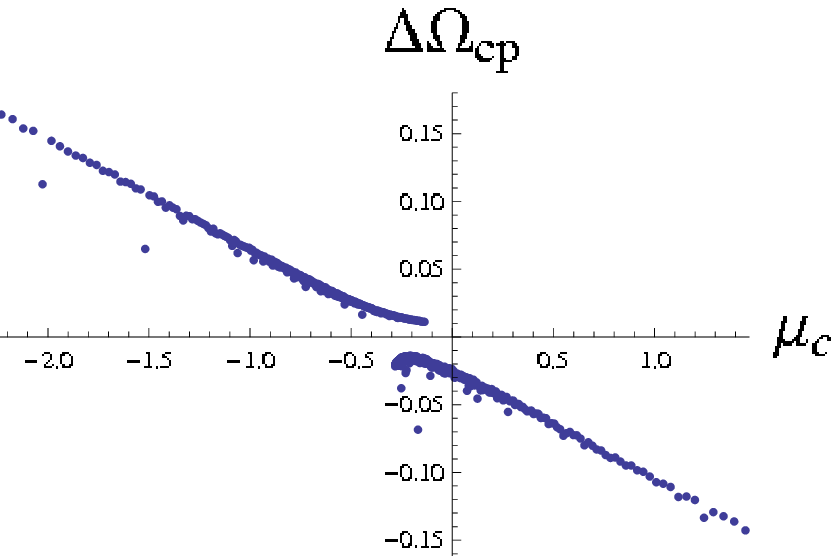}
\caption{$\Delta \Omega_{cp}$ v.s. $\mu_c$ ($L=0.7, T\simeq0.207$)}
\label{Omegacpmuc07_UT075}
\end{center}
\end{minipage}
\end{center}
\end{figure}

\begin{itemize}
\item  For small $L$ ($<0.7$ with $T\simeq0.169$), we find that $\Delta\Omega_{cv}$, $\Delta\Omega_{cp}$, $\Delta\Omega_{pv}$ are all positive, and $\mu_{c}$, $\mu_{p}$ are negative. Thus, the vacuum phase is the most favored. Also from the sign of the susceptibility derived from ${\partial \mu_c(n_b,T) \over \partial n_b}$ and ${\partial \mu_p(n_b,T) \over \partial n_b}$, the cusp and parallel phases are thermodynamically unstable.

\item As $L$ is rising, there is a crossover (starting around  $L=0.7$ with $T\simeq0.169$) and a new branch for both $\Delta\Omega_{cv}$ and $\Delta\Omega_{cp}$ appears as in the confined phase but not for $\Delta\Omega_{pv}$. The new branch has negative $\Delta\Omega_{cv}$ and $\Delta\Omega_{cp}$ so it suggests that the cusp phase is thermodynamically favored. Moreover, ${\partial \mu_c(n_b,T) \over \partial n_b}$ is positive for the new branch so that the cusp phase is also thermodynamically stable. On the other hand, there is no corresponding new branch for ${\partial \mu_p(n_b,T) \over \partial n_b}$ so it implies a thermodynamical instability of the parallel configuration.

\item As $L$ goes higher, we find that there appears no new phase beyond crossover, instead there exists a critical value of $L$ beyond which there is no physical cusp configuration for finite $n_b$. This critical value $L_c$ is $T$-dependent, for example, $L_c\approx 1.28$ for $T\simeq0.131$, $L_c\approx 0.99$ for $T\simeq0.169$, and $L_c\approx 0.82$ for $T\simeq0.207$. Besides, the new branch will tend to be thermodynamically unfavored and the chemical potential of new branch will tend to be negative, although the susceptibility is still positive. This feature is very different from the confined phase where the crossover will end with a dominant stable phase after some value of $L$.

\item For higher T, for example $L=0.7$ with $T\simeq0.207$, we find the similar crossover as for $T\simeq0.169$ case, however, we find that $\Delta\Omega_{cv}$ turns to be positive so that the vacuum phase is now the most favored. For more detailed feature, see Fig.\ref{Omegacvmuc07_UT05}-\ref{Omegacpmuc07_UT075}. Also, as can be seen above, the critical value $L_c$ decreases as $T$ increases.

\item Similar to the confined phase, (though not shown here) there is also a Hagedorn-like phase transition during crossover with a limiting baryonic current density.

\end{itemize}

\begin{figure}[b]
\begin{center}
\begin{minipage}{7.5cm}
\begin{center}\hspace*{-4ex}
\includegraphics[width=6cm]{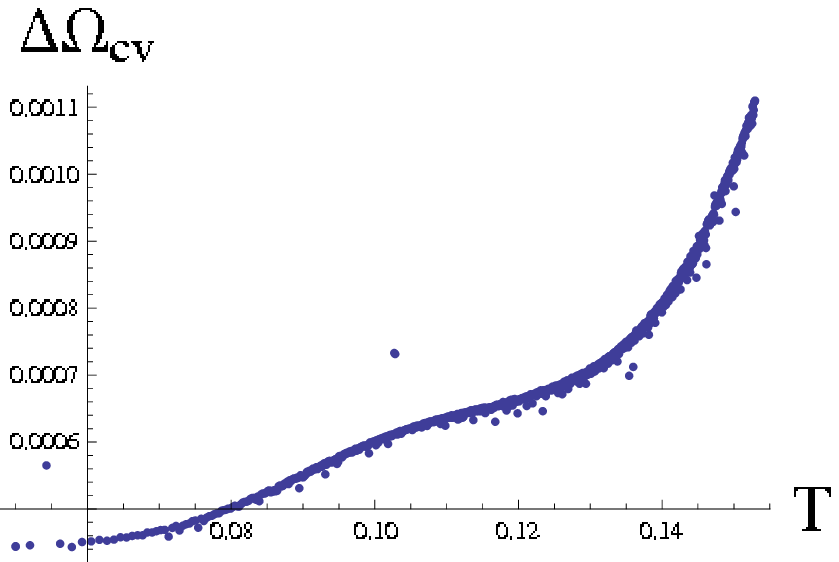}
\caption{$\Delta \Omega_{cv}$ v.s. $T$ ($L=0.7$, $n_{b}=0.005$)}
\label{OmegacvT0005}
\end{center}
\end{minipage}
\hspace{4ex}
\begin{minipage}{7.5cm}
\begin{center}
\includegraphics[width=6cm]{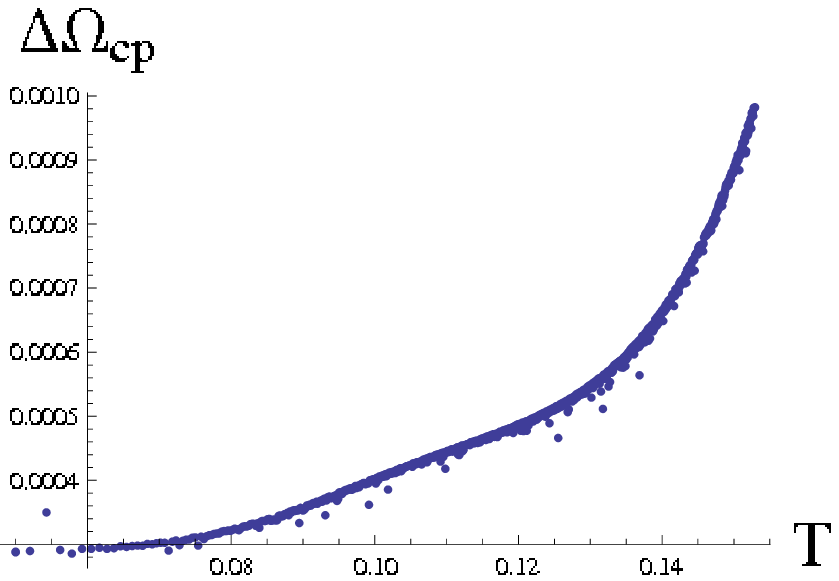}
\caption{$\Omega_{cp}$ v.s. $T$ ($L=0.7$, $n_{b}=0.005$)}
\label{OmegacpT0005}
\end{center}
\end{minipage}
\end{center}
\end{figure}

\begin{figure}[t]
\begin{center}
\begin{minipage}{7.5cm}
\begin{center}
\includegraphics[width=6cm]{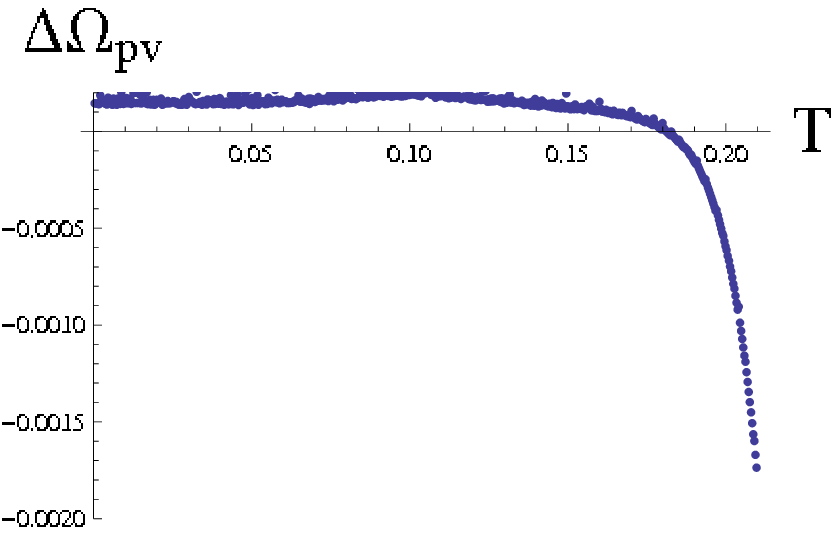}
\caption{$\Delta\Omega_{pv}$ v.s. $T$ ($L=0.7$, $n_{b}=0.005$)}
\label{OmegapvT0005}
\end{center}
\end{minipage}
\hspace{4ex}
\begin{minipage}{7.5cm}
\begin{center}
\includegraphics[width=6cm]{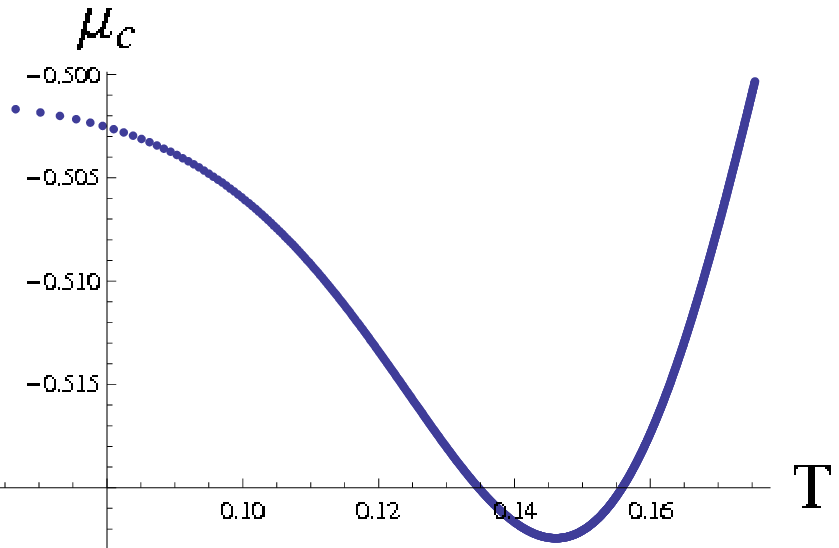}
\caption{$\mu_{c}$ v.s. $T$ ($L=0.7$, $n_{b}=0.05$)}
\label{mucT005}
\end{center}
\end{minipage}
\end{center}
\end{figure}

\begin{figure}[t]
\begin{center}
\begin{minipage}{7.5cm}
\begin{center}
\includegraphics[width=6cm]{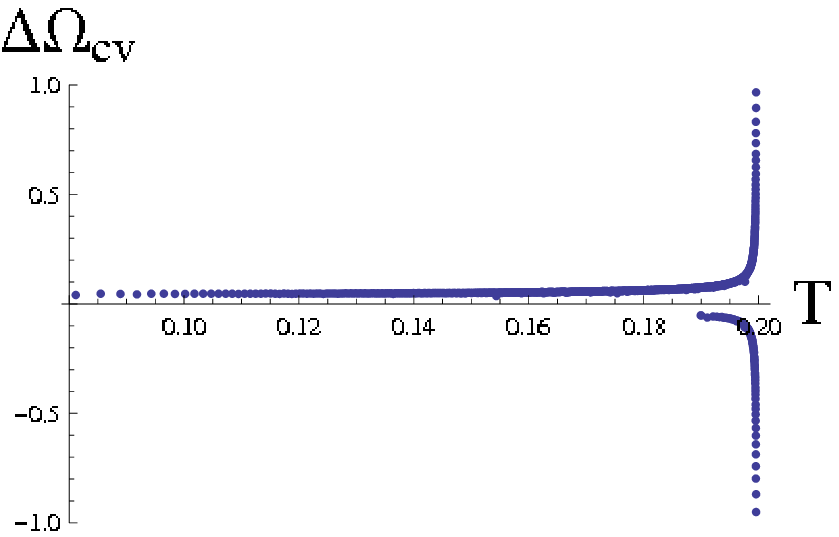}
\caption{$\Omega_{cv}$ v.s. $T$ ($L=0.7$, $n_{b}=0.09$)}
\label{OmegacvT009}
\end{center}
\end{minipage}
\hspace{4ex}
\begin{minipage}{7.5cm}
\begin{center}
\includegraphics[width=6cm]{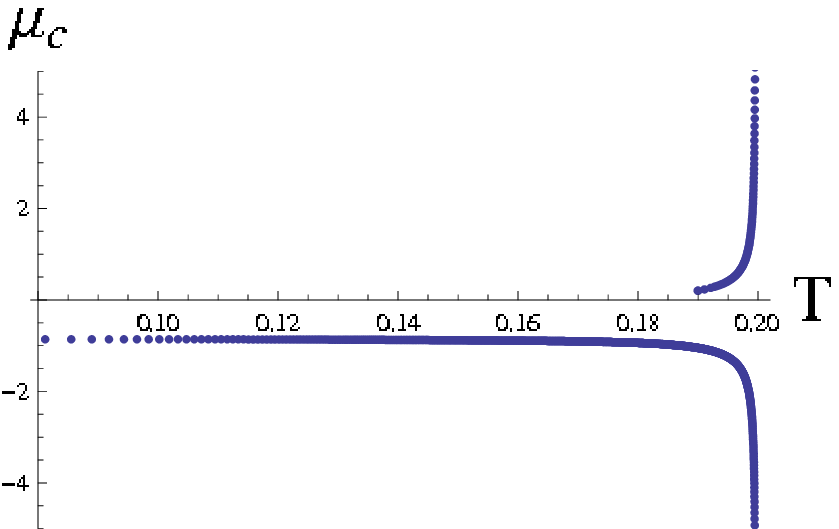}
\caption{$\mu_{c}$ v.s. $T$ ($L=0.7$, $n_{b}=0.09$)}
\label{mucT009}
\end{center}
\end{minipage}
\end{center}
\end{figure}

\subsection*{Fixed $n_{b}$}

  It is interesting to observe the temperature dependence for the fixed baryonic current density. The main features we will summarize in the following:

\begin{itemize}
\item For small $n_{b}$ (e.g. $n_{b}=0.005$), the cusp phase is thermodynamically unfavored. However, $\Delta\Omega_{cv}$ is quite small so that the topologically charged membrane has quite a chance to sustain itself in the vacuum (See Fig.\ref{OmegacvT0005}-\ref{OmegacpT0005}). So is the parallel configuration.

It is also interesting to see in Fig.\ref{OmegapvT0005} there is a critical temperature above which the parallel phase can dominate over the vacuum one, and the chiral symmetry is restored.

\item The thermodynamic relation $-\frac{\mu}{T}=(\frac{\partial S}{\partial n})_{V,U fixed}$ relates the chemical potential to the entropy change, where $U$, $V$ are the internal energy and volume respectively. Therefore, negative $\mu$ yields positive entropy changing rate, and implies the instability.

  The chemical potential $\mu_c$ in Fig.\ref{mucT005} reaches its local minimum at some temperature beyond which it grows but never becomes positive for a physical cusp configuration. That means, the above local minimum of $\mu_c$ yields largest entropy changing rate, and relatively unstable at some specific temperature. On the other hand, when $n_{b}>0.08$ (see Fig.\ref{mucT009}),  the chemical potential is either monotonically decreasing (old branch) or increasing (new branch), so there is no such a special temperature.

\item For $n_{b}$ larger than $0.08$, a crossover phase appears as shown in Fig.\ref{OmegacvT009}-\ref{mucT009} with a new thermodynamically favored and stable cusp branch as seen from $\Delta\Omega_{cv}<0$, $\Delta\Omega_{cp}<0$ and $\mu_{c}>0$. This new branch is also found in the the fixed $T$ case.

   Moreover, in Fig.\ref{mucT009} there is a limiting temperature where $\frac{d\mu}{dT}$ diverges. This is again of Hagedorn-like behavior and implies the meson dissociation as discussed before. This deconfinement temperature found here is very close to the chiral symmetry restoration mentioned above \footnote{The deconfinement phase of finite temperature Sakai-Sugimoto model was first discussed in \cite{Aharony:2006da} as $L$ is varied, however, there is no Hagedorn-like behavior shown as done here.}.

\end{itemize}

\section{Conclusion}

    In this paper we investigate the QCD vacuum with non-trivial topological charge distribution from its holographic dual theory.  We find that there exists thermodynamically favored and stable phase of mesons in the presence of the topologically charged membranes in both low and high temperature regimes. This provides the supporting evidence for the observation of the same topological charge structure in the lattice QCD calculation of \cite{horvath}. we hope our work will inspire more searches for this new vacuum in the lattice calculation, and further estimation of its effect on the quark-gluon plasma created in the heavy-ion collider.    Moreover, from the Chern-Simon term of the probe D8-\AD8-branes, we find that the topologically charged membranes play the role of baryo-axion, which couples to the baryonic current density. This is a kind of holographic realization of the baryo-axion proposed in \cite{dvali}.  We have also shown that the Lorentzian configuration of our topologically charged membrane is dynamically stable, it is interesting to study the modification of the meson and baryon spectra in this new QCD vacuum.

    We should emphasize that our D6-brane fluxes used as the holographic topologically charged membranes are quite different from the ones for $\theta$-vacuum proposed in \cite{witten98}. Our D6-brane fluxes are sourced by the opposite charged sign-coherent sheet, however, the ones in \cite{witten98} are not. It is interesting to point out that the D6-brane fluxes of \cite{witten98} will couple to $A_U$ via Chern-Simon term of D8-\AD8 branes. The lowest KK mode of the gauge field $A_U$ is identified as pion \cite{ss1}, however, this source term will not induce any homogeneous chiral condensate since there is no $F_{U4}$ term in the DBI action of D8-\AD8, which is the same reason for the masslessness of the pion. Instead, the $F_{Ui}, i=1,2,3$ in the DBI action of D8-\AD8 along with the Chern-Simon source term could induce inhomogeneous chiral condensate characterized by $A_U(x^i)$ and the non-trivial $\theta$-vacuum. Further study is needed to make sure if the configuration is thermodynamically favored or not. This is in contrast to the recent proposal for the mass deformation of the Sakai-Sugimoto model \cite{massd}.

    Besides, our results show that the thermodynamics of the topologically charged membranes strongly depends on the value of asymptotic separation of D8-\AD8, and a crossover happens around some critical asymptotic separations. This critical behavior is quite similar to the one for the holographic entanglement where the authors in \cite{Klebanov:2007ws} argue this is related to the Hagedorn spectrum of non-interacting bound states. This is in accordance with our interpretation of the Hagedorn-like meson dissociation phase transition characterized by the limiting baryonic current density or temperature. Despite that, we are not sure why such a critical asymptotic separation exists in our case, it may deserve further study. Also it is important to understand the meaning of the asymptotic separation in the field theory side.

    Finally, we find there is a sign change of chemical potential as we vary the asymptotic separation of D8-\AD8 around the crossover which also happens in the BCS-BEC crossover of some condensed matter system. This analogue motivates the further study to understand the microscopic relation between the topologically charged membrane and the baryonic current density and see if there are corresponding BCS and BEC phases.

\section*{Acknowledgements}
   We would like to thank Hsien-Chung Kao for discussions, especially the help on the numerical implentation, and Eiji Nakano and Takayuki Hirayama for useful comments. We are also appreciated for Dan Tomino's participation at the early stage of this project.  This work was supported by Taiwan's NSC grant 96-2112-M-003-014 and partly by NCTS.


 \end{document}